\documentclass[a4paper,12pt]{article}

\usepackage{ifpdf}

\newif\ifpdf
\ifx\pdfoutput\undefined
  \pdffalse
\else
  \pdfoutput=1
  \pdftrue
\fi

\RequirePackage{xspace} %
\RequirePackage{subfigure} %
\RequirePackage[centertags]{amsmath} %
\RequirePackage{amssymb}
\RequirePackage{wrapfig} %
\RequirePackage{calc} %
\RequirePackage{ifthen}
\RequirePackage{tabularx} %
\RequirePackage{flafter} %
\RequirePackage{fancyhdr} %

\ifpdf
  \RequirePackage[pdftex]{color}%
  \RequirePackage{colortbl}%
  \RequirePackage{array}%
  \RequirePackage[pdftex]{graphicx}

  \RequirePackage[ pdftex, plainpages = false, pdfpagelabels,
                 pdfpagelayout = useoutlines,
                 bookmarks,
                 breaklinks = true,
                 linktocpage,
                 pagebackref,                      
                 colorlinks = true,
                 linkcolor = blue,
                 urlcolor  = blue,
                 citecolor = blue,
                 anchorcolor = blue,
                 hyperindex = true,
                 hyperfigures
                 ]{hyperref}

\else
  \RequirePackage{color}
  \RequirePackage{colortbl}
   \RequirePackage{array}
  \RequirePackage[dvips]{graphicx}
  \RequirePackage{hyperref}
  \usepackage{rotating}
\fi

\usepackage{makeidx} 
\usepackage{setspace} 
\usepackage{rotating} 
\usepackage{ecltree}
\usepackage{epic}
\usepackage{supertabular}  
\usepackage{color}
\usepackage{exscale}
\usepackage{fontenc}
\usepackage{ifthen}
\usepackage{latexsym}
\usepackage{makeidx}
\usepackage{syntonly}
\usepackage{inputenc}
\usepackage{graphicx}
\usepackage{setspace}
\usepackage{caption2}
\usepackage[english]{babel}
\usepackage[square, comma,numbers,sort&compress]{natbib}
\usepackage{hypernat}
\usepackage{boxedminipage}
\usepackage{framed}
\usepackage{longtable}
\usepackage[all]{hypcap}    
\usepackage{algorithm2e}
\usepackage{algorithmic}
\usepackage{lscape}
\usepackage{pdflscape}
\usepackage[T1]{fontenc}
\usepackage{microtype}
\usepackage{mathtools}

\newcommand{\note}[1]{\marginpar[left]{\singlespace \tiny #1}}
\newcommand{\etal}     {{\it et al.}}

\renewcommand{\sectionmark}[1]%
      {\markright{\thesection\ #1}} 

\renewcommand{\note}[1]{}

\newcommand{\CIF}     {\centering \includegraphics[width=2.7in]} %
\newcommand{\Hs}      {\hspace{-0.5cm}} %

\onehalfspace 

\DisableLigatures[f]{encoding = *, family = * }

\begin{document}
\begin{center}
{\Large Further validation to the variational method to obtain flow relations for generalized
Newtonian fluids}
\par\end{center}{\Large \par}

\begin{center}
Taha Sochi
\par\end{center}

\begin{center}
{\scriptsize University College London, Department of Physics \& Astronomy, Gower Street, London,
WC1E 6BT \\ Email: t.sochi@ucl.ac.uk.}
\par\end{center}

\begin{abstract}
\noindent We continue our investigation to the use of the variational method to derive flow
relations for generalized Newtonian fluids in confined geometries. While in the previous
investigations we used the straight circular tube geometry with eight fluid rheological models to
demonstrate and establish the variational method, the focus here is on the plane long thin slit
geometry using those eight rheological models, namely: Newtonian, power law, Ree-Eyring, Carreau,
Cross, Casson, Bingham and Herschel-Bulkley. We demonstrate how the variational principle based on
minimizing the total stress in the flow conduit can be used to derive analytical expressions, which
are previously derived by other methods, or used in conjunction with numerical procedures to obtain
numerical solutions which are virtually identical to the solutions obtained previously from well
established methods of fluid dynamics. In this regard, we use the method of
Weissenberg-Rabinowitsch-Mooney-Schofield (WRMS), with our adaptation from the circular pipe
geometry to the long thin slit geometry, to derive analytical formulae for the eight types of fluid
where these derived formulae are used for comparison and validation of the variational formulae and
numerical solutions. Although some examples may be of little value, the optimization principle
which the variational method is based upon has a significant theoretical value as it reveals the
tendency of the flow system to assume a configuration that minimizes the total stress. Our proposal
also offers a new methodology to tackle common problems in fluid dynamics and rheology.

\vspace{0.3cm}


\par\end{abstract}

\begin{center}

\par\end{center}

\section{Introduction} \label{Introduction}

The flow of Newtonian and non-Newtonian fluids in various confined geometries, such as tubes and
slits, is commonplace in many natural and technological systems. Hence, many methods have been
proposed and developed to solve the flow problems in such geometries applying different physical
principles and employing a diverse collection of analytical, empirical and numerical techniques.
These methods range from employing the first principles of fluid dynamics which are based on the
rules of classical mechanics to more specialized techniques such as the use of
Weissenberg-Rabinowitsch-Mooney-Schofield relation or one of the Navier-Stokes adaptations
\cite{Skellandbook1967, BirdbookAH1987}.

One of the elegant mathematical branches that is regularly employed in the physical sciences is the
calculus of variation which is based on optimizing functionals that describe certain physical
phenomena. The variational method is widely used in many disciplines of theoretical and applied
sciences, such as quantum mechanics and statistical physics, as well as many fields of engineering.
Apart from its mathematical beauty, the method has a big advantage over many other competing
methods by giving an insight into the investigated phenomena. The method does not only solve the
problem formally and hence provides a mathematical solution but it also reveals the Nature habits
and its inclination to economize or lavish on one of the involved physical attributes or the other
such as time, speed, entropy and energy. Some of the well known examples that are based on the
variational principle or derived from the variational method are the Fermat principle of least time
and the curve of fastest descent (brachistochrone). These examples, among many other variational
examples, have played a significant role in the development of the modern natural sciences and
mathematical methods.

In reference \cite{SochiVariational2013} we made an attempt to exploit the variational method to
obtain analytic or numeric relations for the flow of generalized Newtonian fluids in confined
geometries where we postulated that the flow profile in a flow conduit will adjust itself to
minimize the total stress. This attempt was later \cite{SochiVarNonNewt2014} extended to include
more types of non-Newtonian fluids. In the above references, the flow of eight fluid models
(Newtonian, power law, Bingham, Herschel-Bulkley, Carreau, Cross, Ree-Eyring and Casson) in
straight cylindrical tubes was investigated analytically and/or numerically with some of these
models confirming the stated variational hypothesis while others, due to mathematical difficulties
or limitation of the underlying principle, demonstrated behavioral trends that are consistent with
the variational hypothesis.

No mathematically rigorous proof was gives in \cite{SochiVariational2013, SochiVarNonNewt2014} to
establish the proposed variational method that is based on minimizing the total stress in its
generality. Furthermore, we do not make any attempt here to present such a proof. However, in the
present paper we make an attempt to consolidate our previous proposal and findings by giving more
examples, this time from the slit geometry rather than the tube geometry, to validate the use of
the variational principle in deriving flow relations in confined geometries for generalized
Newtonian fluids.

The plan for this paper is as follow. In the next section \S\ \ref{Method} we present the general
formulation of the variational method as applied to the long thin slits and derive the main
variational equation that will be used in obtaining the flow relations for the generalized
Newtonian fluids. In section \S\ \ref{NonViscoplastic} we apply and validate the variational method
for five types of non-viscoplastic fluids, namely: Newtonian, power law, Ree-Eyring, Carreau and
Cross; while in section \S\ \ref{Viscoplastic} we apply and validate the method for three types of
viscoplastic fluids, namely: Casson, Bingham and Herschel-Bulkley. We separate the viscoplastic
from the non-viscoplastic because the variational method strictly applies only to non-viscoplastic
fluids, and hence its use with viscoplastic fluids is an approximation which is valid and good only
when the yield stress value of these fluids is low and hence the departure from fluidity is minor.
In the validation of both non-viscoplastic and viscoplastic types we use the aforementioned WRMS
method where we compare the variational solutions to the analytical solutions obtained from the
WRMS method where analytical formulae for the eight types of fluid are derived in the Appendix (\S\
\ref{Appendix}). The paper in ended in section \S\ \ref{Conclusions} where general discussion and
conclusions about the paper, its objectives and achievements are presented.

\section{Method}\label{Method}

The rheological behavior of generalized Newtonian fluids in one dimensional shear flow is described
by the following constitutive relation

\begin{equation}
\tau=\mu\gamma
\end{equation}
where $\tau$ is the shear stress, $\gamma$ is the rate of shear strain, and $\mu$ is the shear
viscosity which is normally a function of the contemporary rate of shear strain but not of the
deformation history although it may also be a function of other physical parameters such as
temperature and pressure. The latter parameters are not considered in the present investigation as
we assume a static physical setting (i.e. isothermal, isobaric, etc.) apart from the purely
kinematical aspects of the deformation process that is necessary to initiate and sustain the flow.

\begin{figure}[!h]
\centering{}
\includegraphics
[scale=0.75] {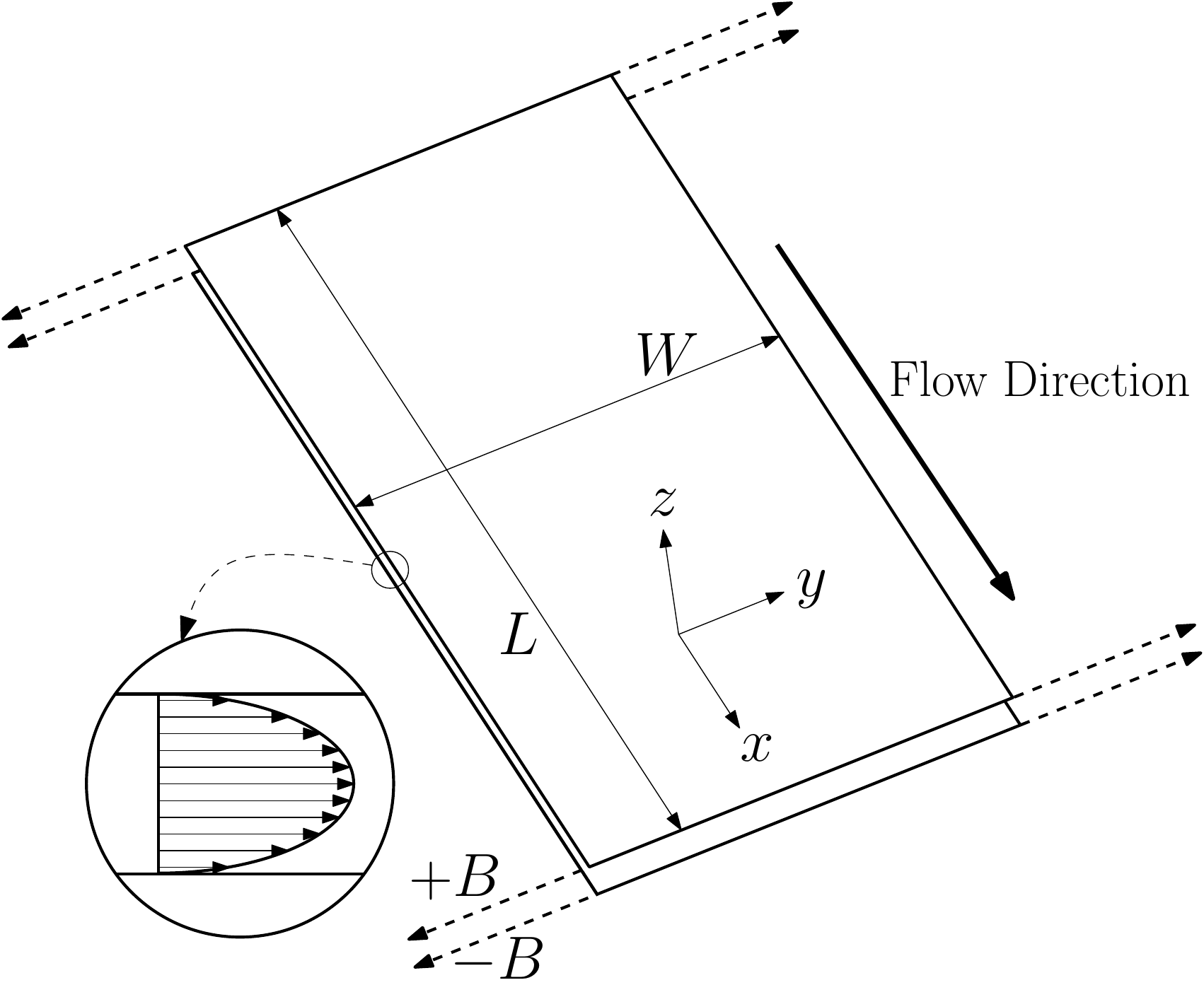} \caption{Schematic drawing of the slit geometry which is used in the
present investigation.} \label{SlitPlot}
\end{figure}

In the following we use the slit geometry, depicted in Figure \ref{SlitPlot}, as our flow apparatus
where $2B$ is the slit thickness, $L$ is the length of the slit across which a pressure drop
$\Delta p$ is imposed, and $W$ is the part of the slit width that is under consideration although
for the purpose of eliminating lateral edge effects we assume that the total width of the slit is
much larger than the considered part $W$. We choose our coordinates system so that the slit
smallest dimension is being positioned symmetrically with respect to the plane $z=0$.

For the slit geometry of Figure \ref{SlitPlot} the total stress is given by

\begin{equation}
\tau_{t}=\int_{\tau_{-B}}^{\tau_{+B}}d\tau=\int_{-B}^{+B}\frac{d\tau}{dz}dz=\int_{-B}^{+B}\frac{d}{dz}\left(\mu\gamma\right)dz=\int_{-B}^{+B}\left(\gamma\frac{d\mu}{dz}+\mu\frac{d\gamma}{dz}\right)dz\label{totalStress}
\end{equation}
where $\tau_{t}$ is the total stress, and $\tau_{\pm B}$ is the shear stress at the slit walls
corresponding to $z=\pm B$.

The total stress, as given by Equation \ref{totalStress}, can be minimized by applying the
Euler-Lagrange variational principle which, in one of its forms, is given by

\begin{equation}\label{EuLa}
\frac{d}{dx}\left(f-y'\frac{\partial f}{\partial y'}\right)-\frac{\partial f}{\partial x}=0
\end{equation}
where the symbols corresponding to our problem statement are defined as

\begin{equation}
x\equiv z,\,\,\,\,\, y\equiv\gamma,\,\,\,\,\, f\equiv\gamma\frac{d\mu}{dz}+\mu\frac{d\gamma}{dz},\,\,\,\mathrm{\,\, and}\,\,\,\,\,\frac{\partial f}{\partial y'}\equiv\frac{\partial}{\partial\gamma'}\left(\gamma\frac{d\mu}{dz}+\mu\frac{d\gamma}{dz}\right)=\mu
\end{equation}

On substituting these symbols into Equation \ref{EuLa} the following equation is obtained

\begin{equation}\label{prev}
\frac{d}{dz}\left(\gamma\frac{d\mu}{dz}+\mu\frac{d\gamma}{dz}-\mu\frac{d\gamma}{dz}\right)-\frac{\partial}{\partial
z}\left(\gamma\frac{d\mu}{dz}+\mu\frac{d\gamma}{dz}\right)=0
\end{equation}
Considering the fact that for the considered flow systems

\begin{equation}
\gamma\frac{d\mu}{dz}+\mu\frac{d\gamma}{dz}=G
\end{equation}
where $G$ is a constant, it can be shown that Equation \ref{prev} can be reduced to two independent
variational forms

\begin{equation}\label{MainVar1}
\frac{d}{d z}\left(\gamma\frac{d\mu}{dz}\right)=0
\end{equation}
and

\begin{equation}\label{MainEq}
\frac{d}{d z}\left(\mu\frac{d\gamma}{dz}\right)=0
\end{equation}

In the following two sections we use the second form where we outline the application of the
variational method, as summarized in Equation \ref{MainEq}, to validate and demonstrate the use of
the variational method to obtain flow relations correlating the volumetric flow rate $Q$ through
the slit to the pressure drop $\Delta p$ across the slit length $L$ for generalized Newtonian
fluids. The plan is that we derive fully analytical expressions when this is viable, as in the case
of Newtonian and power law fluids, and partly analytical solutions when the former is not viable,
e.g. for Ree-Eyring and Herschel-Bulkley fluids. In the latter case, the solution is obtained
numerically in its final stages, following a variationally based derivation, by using numerical
integration and simple numerical solvers.

In this investigation we assume a laminar, incompressible, time-independent, fully-developed,
isothermal flow where entry and exit edge effects are negligible. We also assume negligible body
forces and a blunt flow speed profile with a no-shear stationary region at the profile center plane
which is consistent with the considered type of fluids and flow conditions, i.e. viscous
generalized Newtonian fluids in a pressure-driven laminar flow. As for the plane slit geometry, we
assume, following what is stated in the literature, a long thin slit with $B\ll W$ and $B\ll L$
although we believe that some of these conditions are redundant according to our own statement and
problem settings. We also assume that the slit is rigid and uniform in shape and size, that is its
walls are not made of deformable materials, such as elastic or viscoelastic, and the slit does not
experience an abrupt or gradual change in $B$.

\section{Non-Viscoplastic Fluids}\label{NonViscoplastic}

For non-viscoplastic fluids, the variational principle strictly applies. In this section we apply
the variational method to five non-viscoplastic fluids and compare the variational solutions to the
analytical solutions obtained from the WRMS method. These fluids are: Newtonian, power law,
Ree-Eyring, Carreau and Cross. All these models have analytical solutions that can be obtained from
various traditional methods of fluid dynamics which are not based on the variational principle.
Hence the agreement between the solutions obtained from the traditional methods with the solutions
obtained from the variational method will validate and vindicate the variational approach.

As indicated earlier, to derive non-variational analytical relations we use a method similar to the
one ascribed to Weissenberg, Rabinowitsch, Mooney, and Schofield \cite{Skellandbook1967} for the
flow of generalized Newtonian fluids in uniform tubes with circular cross sections, where we adapt
and apply the procedure to long thin slits, and hence we label this method with WRMS to abbreviate
the names of its originators. The WRMS method is fully explained and applied in the Appendix (\S\
\ref{Appendix}) to derive analytical relations to all the eight fluid models that are used in the
present paper. For some of these fluids full analytical solutions from the variational principle
are obtained and hence a direct comparison between the analytical expressions obtained from the two
methods can be made, while for other fluids a mixed analytical and numerical procedure is employed
to obtain numerical solutions from the variational principle and hence a representative sample of
numerical solutions from both methods is presented for comparison and validation, as will be
clarified and demonstrated in the following subsections.

\subsection{Newtonian}\label{NewtonianSec}

The viscosity of Newtonian fluids is constant, that is

\begin{equation}
\mu=\mu_{o}\end
{equation}
and therefore Equation \ref{MainEq} becomes

\begin{equation}
\frac{d}{d z}\left(\mu_{o}\frac{d\gamma}{dz}\right)=0
\end{equation}
On performing the outer integration we obtain

\begin{equation}
\mu_{o}\frac{d\gamma}{dz}=A
\end{equation}
where $A$ is the constant of integration. On performing the inner integration we obtain

\begin{equation}
\gamma=\frac{A}{\mu_{o}}z+D
\end{equation}
where $D$ is a second constant of integration. Now from the no-shear condition at the slit center
plane $z=0$, $D$ can be determined, that is

\begin{equation}
\gamma\left(z=0\right)=0\,\,\,\,\,\,\,\,\,\Rightarrow\,\,\,\,\,\,\,\, D=0
\end{equation}
Similarly, from the no-slip boundary condition \cite{SochiSlip2011} at $z=\pm B$ which controls the
wall shear stress we determine $A$, i.e.

\begin{equation}\label{tauBeq}
\tau_{\pm B}=\frac{F_{\perp}}{\sigma_{\parallel}}=\frac{2BW\Delta p}{2WL}=\frac{B \Delta p}{L}
\end{equation}
where $\tau_{\pm B}$ is the shear stress at the slit walls, $F_{\perp}$ is the flow driving force
which is normal to the slit cross section in the flow direction, and $\sigma_{\parallel}$ is the
area of the slit walls which is tangential to the flow direction. Hence

\begin{equation}
\gamma\left(z=\pm B\right)=\frac{\tau_{\pm B}}{\mu_{o}}=\frac{B\Delta
p}{\mu_{o}L}=\frac{AB}{\mu_{o}}\,\,\,\,\,\,\,\,\,\Rightarrow\,\,\,\,\,\,\,\, A=\frac{\Delta p}{L}
\end{equation}
Therefore

\begin{equation}
\gamma\left(z\right)=\frac{\Delta p}{\mu_{o}L}z\label{NewtonianGamR}
\end{equation}
On integrating the rate of shear strain with respect to $z$, the standard parabolic speed profile
is obtained, that is

\begin{equation}
v\left(z\right)=\int dv=\int\frac{dv}{dz}dz=\int-\gamma dz=-\int\frac{\Delta
p}{\mu_{o}L}zdz=-\frac{\Delta p}{2\mu_{o}L}z^{2}+E
\end{equation}
where $v(z)$ is the fluid speed at $z$ in the $x$ direction and $E$ is another constant of
integration which can be determined from the no-slip at the wall boundary condition, that is

\begin{equation}
v\left(z=\pm B\right)=0\,\,\,\,\,\,\,\,\Rightarrow E=\frac{\Delta p}{2\mu_{o}L}B^{2}
\end{equation}
i.e.

\begin{equation}
v\left(z\right)=\frac{\Delta p}{2\mu_{o}L}\left(B^{2}-z^{2}\right)
\end{equation}
The volumetric flow rate is then obtained by integrating the
flow speed profile over the slit cross sectional area in the $z$ direction, that is

\begin{equation}
Q=\int_{-B}^{+B}v\, Wdz=\frac{2W\Delta
p}{2\mu_{o}L}\int_{0}^{B}\left(B^{2}-z^{2}\right)dz=\frac{W\Delta
p}{\mu_{o}L}\left[B^{2}z-\frac{z^{3}}{3}\right]_{0}^{B}\label{NewtonianQP}
\end{equation}
that is

\begin{equation}
Q=\frac{2WB^{3}\Delta p}{3\mu_{o}L}
\end{equation}
which is the well known volumetric flow rate formula for the flow of Newtonian fluids in a plane
long thin slit as obtained by other methods which are not based on the variational principle. This
formula is derived in the Appendix (\S\ \ref{Appendix}, Equation \ref{QeqNewt}) using the WRMS
method. It also can be found in several classic textbooks of fluid mechanics, e.g. Bird \etal\
\cite{BirdbookAH1987} Table 4.5-14 where $\mu_{o}\equiv\mu$ and $\Delta p\equiv P_{0}-P_{L}$.

\subsection{Power Law}\label{PowerLawSec}

The shear dependent viscosity of power law fluids is given by \cite{Skellandbook1967,
BirdbookAH1987, CarreaubookKC1997}

\begin{equation}
\mu=k\gamma^{n-1}
\end{equation}
where $k$ is the power law viscosity consistency coefficient and $n$ is the flow behavior index. On
applying the Euler-Lagrange variational principle (Equation \ref{MainEq}) we obtain

\begin{equation}
\frac{d}{d z}\left(k\gamma^{n-1}\frac{d\gamma}{dz}\right)=0
\end{equation}
On performing the outer integral we obtain

\begin{equation}
k\gamma^{n-1}\frac{d\gamma}{dz}=A
\end{equation}
On separating the two variables in the last equation and integrating both sides we obtain

\begin{equation}
\gamma=\sqrt[n]{\frac{n}{k}\left(Az+D\right)}\label{plGam}
\end{equation}
where $A$ and $D$ are the constants of integration which can be determined from the two limiting
conditions, that is

\begin{equation}
\gamma\left(z=0\right)=0\,\,\,\,\,\,\,\,\,\Rightarrow\,\,\,\,\,\,\,\, D=0
\end{equation}
and

\begin{equation}
\gamma\left(z=B\right)=\sqrt[n]{\frac{\tau_{B}}{k}}=\sqrt[n]{\frac{B\Delta
p}{Lk}}=\sqrt[n]{\frac{n}{k}AB}\,\,\,\,\,\,\,\,\,\Rightarrow\,\,\,\,\,\,\,\, A=\frac{\Delta p}{nL}
\end{equation}
where the first step in the last equation is obtained from the constitutive relation of power law
fluids, i.e.

\begin{equation}
\tau=k\gamma^{n}
\end{equation}
with the substitution $z=B$ in Equation \ref{plGam}. Hence, from Equation \ref{plGam} we obtain

\begin{equation}
\gamma=\sqrt[n]{\frac{\Delta p}{kL}}z^{1/n}\label{PowerLawGamR}
\end{equation}
On integrating the rate of shear strain with respect to $z$, the flow speed profile is determined,
i.e.

\begin{equation}
v\left(z\right)=\int dv=\int\frac{dv}{dz}dz=-\int\gamma dz=-\int\sqrt[n]{\frac{\Delta
p}{kL}}z^{1/n}dz=-\frac{n}{n+1}\sqrt[n]{\frac{\Delta p}{kL}}z^{1+1/n}+E
\end{equation}
where $E$ is another constant of integration which can be determined from the no-slip at the wall
condition, that is

\begin{equation}
v\left(z=B\right)=0\,\,\,\,\,\,\,\, \Rightarrow \,\,\,\,\,\,\,\,
E=\frac{n}{n+1}\sqrt[n]{\frac{\Delta p}{kL}}B^{1+1/n}
\end{equation}
i.e.

\begin{equation}
v\left(z\right)=\frac{n}{n+1}\sqrt[n]{\frac{\Delta p}{kL}}\left(B^{1+1/n}-z^{1+1/n}\right)
\end{equation}
The volumetric flow rate can then be obtained by integrating the flow speed profile with respect to
the cross sectional area in the $z$ direction, that is

\begin{equation}
Q=\int_{-B}^{+B}v\, Wdz=\frac{2Wn}{n+1}\sqrt[n]{\frac{\Delta
p}{kL}}\int_{0}^{B}\left(B^{1+1/n}-z^{1+1/n}\right)dz
\end{equation}

\begin{equation}
=\frac{2Wn}{n+1}\sqrt[n]{\frac{\Delta
p}{kL}}\left[B^{1+1/n}z-\frac{z^{2+1/n}}{2+1/n}\right]_{0}^{B}\label{PowerLawQP}
\end{equation}

\begin{equation}
=\frac{2Wn}{n+1}\sqrt[n]{\frac{\Delta
p}{kL}}\left[B^{2+1/n}-\frac{B^{2+1/n}}{2+1/n}\right]\label{PowerLawQP}
\end{equation}
i.e.

\begin{equation}
Q=\frac{2WB^{2}n}{2n+1}\sqrt[n]{\frac{B\Delta p}{kL}}\label{PowerLawQP}
\end{equation}
which is the well known volumetric flow rate relation for the flow of power law fluids in a long
thin slit as obtained by other non-variational methods. This formula is derived in the Appendix
(\S\ \ref{Appendix}, Equation \ref{QeqPl}) using the WRMS method. It can also be found in textbooks
of fluid mechanics such as Bird \etal\ \cite{BirdbookAH1987} Table 4.2-1 where $k\equiv m$ and
$\Delta p\equiv P_{0}-P_{L}$.

\subsection{Ree-Eyring}

For Ree-Eyring fluids, the constitutive relation between shear stress and rate of strain is given
by \cite{BirdbookAH1987}

\begin{equation}
\tau=\tau_{c}\,\mathrm{asinh}\left(\frac{\mu_{r}\gamma}{\tau_{c}}\right)
\end{equation}
where $\tau_{c}$ is a characteristic shear stress and $\mu_{r}$ is the viscosity at vanishing rate
of strain. Hence, the generalized Newtonian viscosity is given by

\begin{equation}
\mu=\frac{\tau}{\gamma}=\frac{\tau_{c}\,\mathrm{asinh}\left(\frac{\mu_{r}\gamma}{\tau_{c}}\right)}{\gamma}
\end{equation}
On substituting $\mu$ from the last relation into Equation \ref{MainEq} we obtain

\begin{equation}
\frac{d}{d
z}\left(\frac{\tau_{c}\,\mathrm{asinh}\left(\frac{\mu_{r}\gamma}{\tau_{c}}\right)}{\gamma}\frac{d\gamma}{dz}\right)=0
\end{equation}
On integrating once we get

\begin{equation}
\frac{\tau_{c}\,\mathrm{asinh}\left(\frac{\mu_{r}\gamma}{\tau_{c}}\right)}{\gamma}\frac{d\gamma}{dz}=A
\end{equation}
where $A$ is a constant. On separating the two variables and integrating again we obtain

\begin{equation}\label{numInt}
\int\frac{\tau_{c}\,\mathrm{asinh}\left(\frac{\mu_{r}\gamma}{\tau_{c}}\right)}{\gamma}d\gamma=Az
\end{equation}
where the constant of integration $D$ is absorbed within the integral on the left hand side. The
integral on the left hand side of Equation \ref{numInt} when evaluated analytically produces an
expression that involves logarithmic and polylogarithmic functions which when computed produce
significant errors especially in the neighborhood of $z=0$. To solve this problem we used a
numerical integration procedure to evaluate this integral, and hence obtain $A$, numerically using
the boundary condition at the slit wall, that is

\begin{equation}
\gamma\left(z=B\right)\equiv\gamma_{B}=\frac{\tau_{c}}{\mu_{r}}\sinh\left(\frac{\tau_{B}}{\tau_{c}}\right)
\end{equation}
where $\tau_{B}$ is given by Equation \ref{tauBeq}. This was then followed by obtaining $\gamma$ as
a function of $z$ using a bisection numerical solver in conjunction with a numerical integration
procedure based on Equation \ref{numInt}. Due to the fact that the constant of integration, $D$, is
absorbed in the left hand side and a numerical integration procedure was used rather than an
analytical evaluation of the integral on the left hand side of Equation \ref{numInt}, there was no
need for an analytical evaluation of this constant using the boundary condition at the slit center
plane, i.e.
\begin{equation}
\gamma\left(z=0\right)=0
\end{equation}
The numerically obtained $\gamma$ was then integrated numerically with respect to $z$ to obtain the
flow speed as a function of $z$ where the no-slip boundary condition at the wall is used to have an
initial value $v=0$ that is incremented on moving inward from the wall toward the center plane. The
flow speed profile was then integrated numerically with respect to the cross sectional area to
obtain the volumetric flow rate.

To test the validity of the variational method we made extensive comparisons between the solutions
obtained from the variational method to those obtained from the WRMS method using widely varying
ranges of fluid and slit parameters. In Figure \ref{ReeEyringFig} we compare the WRMS analytical
solutions as derived in the Appendix (\S\ \ref{Appendix}, Equation \ref{QeqRE}) with the
variational solutions for two sample cases. As seen, the two methods agree very well which is
typical in all the investigated cases. The minor differences between the solutions of the two
methods can be easily explained by the accumulated errors arising from repetitive use of numerical
integration and numerical solvers in the variational method. The errors as estimated by the
percentage relative difference are typically less than 0.5\% when using reliable numerical
integration schemes with reasonably fine discretization mesh and tiny error margin for the
convergence condition of the numerical solver. This is also true in general for the other types of
fluid that will follow in this section.


\begin{figure}
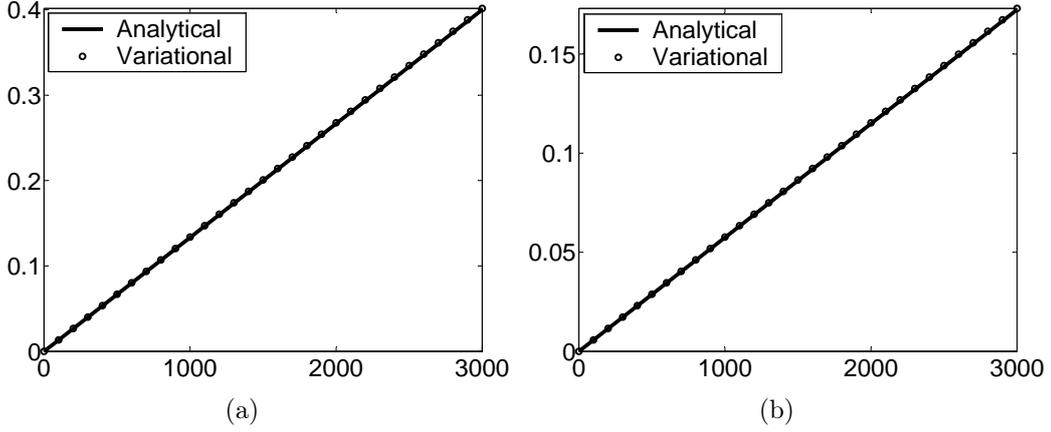

\centering %
\subfigure[]%
{\begin{minipage}[b]{0.5\textwidth} \CIF {g/ReeEyringFig1}
\end{minipage}}
\Hs
\subfigure[]%
{\begin{minipage}[b]{0.5\textwidth} \CIF {g/ReeEyringFig2}
\end{minipage}}
%
\caption{Comparing the WRMS analytical solutions, as given by Equation \ref{QeqRE}, to the
variational solutions for the flow of Ree-Eyring fluids in long thin slits with
(a) $\mu_r=0.005$~Pa.s, $\tau_c=600$~Pa, $B=0.01$~m, $W=1.0$~m and $L=1.0$~m;
and (b) $\mu_r=0.015$~Pa.s, $\tau_c=400$~Pa, $B=0.013$~m, $W=1.0$~m and $L=1.7$~m.
In both sub-figures, the vertical axis represents the volumetric flow rate, $Q$, in m$^3$.s$^{-1}$
while the horizontal axis represents the pressure drop, $\Delta p$, in Pa. The average percentage
relative difference between the WRMS analytical solutions and the variational solutions for these
cases are about 0.38\% and 0.39\% respectively. \label{ReeEyringFig}}
\end{figure}

\subsection{Carreau}

For Carreau fluids, the viscosity is given by \cite{BirdbookAH1987, Sorbiebook1991,
CarreaubookKC1997, Tannerbook2000}

\begin{equation}
\mu=\mu_{i}+\left(\mu_{0}-\mu_{i}\right)\left[1+\lambda^{2}\gamma^{2}\right]^{\left(n-1\right)/2}
\end{equation}
where $\mu_{0}$ is the zero-shear viscosity, $\mu_{i}$ is the infinite-shear viscosity, $\lambda$
is a characteristic time constant, and $n$ is the flow behavior index. On applying the
Euler-Lagrange variational principle (Equation \ref{MainEq}) and following the derivation, as
outlined in the previous subsections, we obtain

\begin{equation}\label{CarMain}
\mu_{i}\gamma+\left(\mu_{0}-\mu_{i}\right)\gamma\,_{2}F_{1}\left(\frac{1}{2},\frac{1-n}{2};\frac{3}{2};-\lambda^{2}\gamma^{2}\right)=Az+D
\end{equation}
where $_{2}F_{1}$ is the hypergeometric function of the given argument with its real part being
used in the computation, and $A$ and $D$ are the constants of integration. From the two boundary
conditions at $z=0$ and $z=B$, $A$ and $D$ can be determined, that is

\begin{equation}
\gamma\left(z=0\right)=0\,\,\,\,\,\,\,\,\,\Rightarrow\,\,\,\,\,\,\,\, D=0
\end{equation}
and

\begin{equation}\label{CarBound2}
\gamma\left(z=B\right)=\gamma_{B}\,\,\,\,\,\,\,\,\,\Rightarrow\,\,\,\,\,\,\,\,\mu_{i}\gamma_{B}+\left(\mu_{0}-\mu_{i}\right)\gamma_{B}\,_{2}F_{1}\left(\frac{1}{2},\frac{1-n}{2};\frac{3}{2};-\lambda^{2}\gamma_{B}^{2}\right)=AB
\end{equation}
where $\gamma_{B}$ is the shear rate at the slit wall. Now, by definition we have

\begin{equation}
\mu_{B}\gamma_{B}=\tau_{B}
\end{equation}
that is

\begin{equation}
\left[\mu_{i}+\left(\mu_{0}-\mu_{i}\right)\left[1+\lambda^{2}\gamma_{B}^{2}\right]^{\left(n-1\right)/2}\right]\gamma_{B}=\frac{B\Delta
p}{L}
\end{equation}
From the last equation, $\gamma_{B}$ can be obtained numerically by a numerical solver, based for
example on a bisection method, and hence from Equation \ref{CarBound2} $A$ is obtained. Equation
\ref{CarMain} can then be solved numerically to obtain the shear rate $\gamma$ as a function of
$z$. This will be followed by integrating $\gamma$ numerically with respect to $z$ to obtain the
speed profile, $v(r)$, where the no-slip boundary condition at the wall is used to have an initial
value $v(z=\pm B)=0$ that is incremented on moving inward toward the center plane. The speed
profile, in its turn, will be integrated numerically with respect to the slit cross sectional area
to obtain the volumetric flow rate $Q$.

\begin{figure}
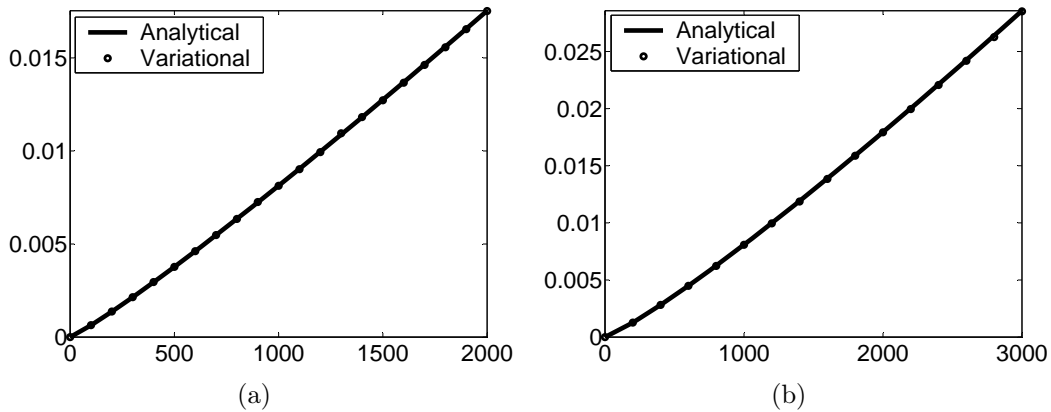

\centering %
\subfigure[]%
{\begin{minipage}[b]{0.5\textwidth} \CIF {g/CarreauFig1}
\end{minipage}}
\Hs
\subfigure[]%
{\begin{minipage}[b]{0.5\textwidth} \CIF {g/CarreauFig2}
\end{minipage}}
%
\caption{Comparing the WRMS analytical solutions, as given by Equation \ref{QeqCar}, to the
variational solutions for the flow of Carreau fluids in long thin slits with
(a) $n=0.9$, $\mu_0=0.13$~Pa.s, $\mu_{i}=0.002$~Pa.s, $\lambda=0.8$~s, $B=0.011$~m, $W=1.0$~m and
$L=1.25$~m;
and (b) $n=0.85$, $\mu_0=0.15$~Pa.s, $\mu_{i}=0.01$~Pa.s, $\lambda=1.65$~s, $B=0.011$~m, $W=1.0$~m
and $L=1.4$~m.
The axes are as in Figure \ref{ReeEyringFig}, while the differences are about 0.21\% and 0.25\%
respectively.
\label{CarreauFig}}
\end{figure}

In Figure \ref{CarreauFig} we present two sample cases for the flow of Carreau fluids in thin slits
where the WRMS analytical solutions, as given by Equation \ref{QeqCar}, are compared to the
variational solutions. Good agreement can be seen in these plots which are typical of the
investigated cases. The main reason for the difference between the WRMS and variational solutions
is the persistent use of numerical solvers and numerical integration in the implementation of the
variational method as well as the use of the hypergeometric function in both methods. The numerical
implementation of this function can cause instability and failure to converge satisfactorily in
some cases.


\subsection{Cross}

For Cross fluids, the viscosity is given by \cite{CarreaubookKC1997, OwensbookP2002}

\begin{equation}
\mu=\mu_{i}+\frac{\mu_{0}-\mu_{i}}{1+\lambda^{m}\gamma^{m}}
\end{equation}
where $\mu_{0}$ is the zero-shear viscosity, $\mu_{i}$ is the infinite-shear viscosity, $\lambda$
is a characteristic time constant, and $m$ is an indicial parameter. Following a similar derivation
method to that outlined in Carreau, we obtain

\begin{equation}\label{CroMain}
\mu_{i}\gamma+\left(\mu_{0}-\mu_{i}\right)\gamma\,_{2}F_{1}\left(1,\frac{1}{m};\frac{m+1}{m};-\lambda^{m}\gamma^{m}\right)=Az
\end{equation}
where

\begin{equation}
A=\frac{\mu_{i}\gamma_{B}+\left(\mu_{0}-\mu_{i}\right)\gamma_{B}\,_{2}F_{1}\left(1,\frac{1}{m};\frac{m+1}{m};-\lambda^{m}\gamma_{B}^{m}\right)}{B}
\end{equation}
with $\gamma_{B}$ being obtained numerically as outlined in Carreau. Equation \ref{CroMain} can
then be solved numerically to obtain the shear rate $\gamma$ as a function of $z$. This is followed
by obtaining $v$ from $\gamma$ and $Q$ from $v$ by using numerical integration, as before.

\begin{figure}
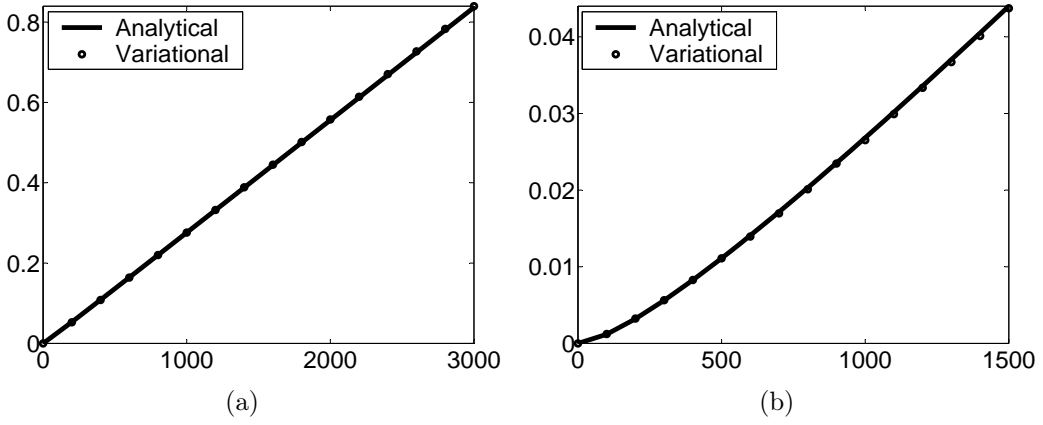

\centering %
\subfigure[]%
{\begin{minipage}[b]{0.5\textwidth} \CIF {g/CrossFig1}
\end{minipage}}
\Hs
\subfigure[]%
{\begin{minipage}[b]{0.5\textwidth} \CIF {g/CrossFig2}
\end{minipage}}
%
\caption{Comparing the WRMS analytical solutions, as given by Equation \ref{QeqCro}, to the
variational solutions for the flow of Cross fluids in long thin slits with
(a) $m=0.65$, $\mu_0=0.032$~Pa.s, $\mu_{i}=0.004$~Pa.s, $\lambda=4.5$~s, $B=0.013$~m, $W=1.0$~m and
$L=1.3$~m;
and (b) $m=0.5$, $\mu_0=0.075$~Pa.s, $\mu_{i}=0.004$~Pa.s, $\lambda=0.8$~s, $B=0.006$~m, $W=1.0$~m
and $L=0.8$~m.
The axes are as in Figure \ref{ReeEyringFig}, while the differences are about 0.29\% and 0.88\%
respectively. \label{CrossFig}}
\end{figure}

In Figure \ref{CrossFig} we present two sample cases for the flow of Cross fluids in thin slits
where we compare the WRMS analytical solutions, as given by Equation \ref{QeqCro}, to the
variational solutions. As seen in these plots, the agreement is good with the main reason for the
departure between the two methods is the persistent use of numerical solvers and numerical
integration in the variational method as well as the use of the hypergeometric function in both
methods, as explained in the case of Carreau.


\section{Viscoplastic Fluids}\label{Viscoplastic}

The yield stress fluids are not strictly subject to the variational principle due to the existence
of a solid non-yield region at the center which does not obey the stress optimization condition and
hence the Euler-Lagrange variational method is not strictly applicable to these fluids. However,
the method provides a good approximation when the value of the yield stress is low so that the
effect of the non-yield region at and around the center plane of the slit on the flow profile is
minor. In the following subsections we apply the variational method to three yield stress fluids
and obtain some solutions from sample cases which are representative of the many cases that were
investigated.

\subsection{Casson}

For Casson fluids, the constitutive relation is given by \cite{BirdbookAH1987, CarreaubookKC1997}

\begin{equation}
\tau=\left[\left(K\gamma\right)^{1/2}+\tau_{o}^{1/2}\right]^{2}
\end{equation}
where $K$ is the viscosity consistency coefficient, and $\tau_{o}$ is the yield stress. Hence

\begin{equation}
\mu=\frac{\tau}{\gamma}=\frac{\left[\left(K\gamma\right)^{1/2}+\tau_{o}^{1/2}\right]^{2}}{\gamma}
\end{equation}
On substituting $\mu$ from the last equation into the main variational relation, as given by
Equation \ref{MainEq}, we obtain

\begin{equation}
\frac{d}{d
z}\left(\frac{\left[\left(K\gamma\right)^{1/2}+\tau_{o}^{1/2}\right]^{2}}{\gamma}\frac{d\gamma}{dz}\right)=0
\end{equation}
On integrating twice with respect to $z$ we obtain

\begin{equation}
K\gamma+4\left(K\tau_{o}\gamma\right)^{1/2}+\tau_{o}\ln\left(\gamma\right)=Az+D\label{CasMain}
\end{equation}
where $A$ and $D$ are constants. Now, from the boundary condition at the slit center plane we have

\begin{equation}
\gamma\left(z=0\right)=0
\end{equation}
so we set $D=0$ to constrain the solution at $z=0$. As for the second boundary condition at the
slit wall, $z=B$, we have

\begin{equation}
K\gamma_{B}+4\left(K\tau_{o}\gamma_{B}\right)^{1/2}+\tau_{o}\ln\left(\gamma_{B}\right)=AB
\end{equation}
where the rate of shear strain at the slit wall, $\gamma_B$, is obtained from applying the
constitutive relation at the wall and hence is given by

\begin{equation}
\gamma_{B}=\frac{\left[\sqrt{\tau_{B}}-\tau_{o}^{1/2}\right]^{2}}{K}
\end{equation}
with the shear stress at the slit wall, $\tau_B$, being given by Equation \ref{tauBeq}. Hence

\begin{equation}
A=\frac{K\gamma_{B}+4\left(K\tau_{o}\gamma_{B}\right)^{1/2}+\tau_{o}\ln\left(\gamma_{B}\right)}{B}
\end{equation}
Equation \ref{CasMain} defines $\gamma$ implicitly in terms of $z$ and hence it is solved
numerically using, for instance, a numerical bisection method to find $\gamma$ as a function of $z$
with avoidance of the very immediate neighborhood of $z=0$ which, as explained earlier, is not
subject to the variational method. This is equivalent to integrating between $\tau_o$ and $\tau_B$,
rather than between 0 and $\tau_B$, in the WRMS method, as employed in the Appendix (\S\
\ref{Appendix}), for the case of Casson, Bingham and Herschel-Bulkley fluids. Although the value of
$z$ that defines the start of the yield region near the center is not known exactly, we already
assumed that the use of the variational method is only legitimate when $\tau_o$ is small and hence
the non-yield plug region is small and hence its effect is minor, so any error from an ambiguity in
the exact limit of the integral near the $z=0$ should be negligible especially at high flow rates
where this region shrinks and hence using a lower limit of the integral at the immediate
neighborhood of $z=0$ will give a more exact definition of the yield region. On obtaining $\gamma$
numerically, $v$ and $Q$ can be obtained successively by numerical integration, as before.

\begin{figure}
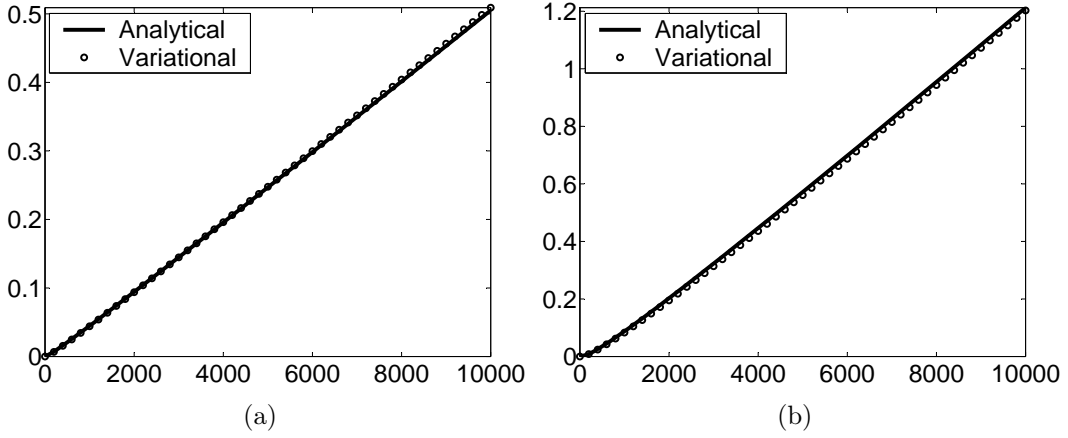

\centering %
\subfigure[]%
{\begin{minipage}[b]{0.5\textwidth} \CIF {g/CassonFig1}
\end{minipage}}
\Hs
\subfigure[]%
{\begin{minipage}[b]{0.5\textwidth} \CIF {g/CassonFig2}
\end{minipage}}
%
\caption{Comparing the WRMS analytical solutions, as given by Equation \ref{QeqCas}, to the
variational solutions for the flow of Casson fluids in long thin slits with
(a) $K=0.025$~Pa.s, $\tau_o=0.1$~Pa, $B=0.01$~m, $W=1.0$~m and $L=0.5$~m;
and (b) $K=0.05$~Pa.s, $\tau_o=0.5$~Pa, $B=0.025$~m, $W=1.0$~m and $L=1.5$~m.
The axes are as in Figure \ref{ReeEyringFig}, while the differences are about 0.57\% and 2.84\%
respectively. \label{CassonFig}}
\end{figure}

In Figure \ref{CassonFig} we present two sample cases for the flow of Casson fluids in thin slits
where we compare the WRMS analytical solutions, as given by Equation \ref{QeqCas}, with the
variational solutions. As seen, the agreement is reasonably good considering that the variational
method is just an approximation and hence it is not supposed to produce identical results to the
analytically derived solutions. The two plots also indicate that the approximation is worsened as
the value of the yield stress increases, resulting in the increase of the effect of the non-yield
region at the center plane of the slit which is not subject to the variational principle, and hence
more deviation between the two methods is observed.


\subsection{Bingham}

For Bingham fluids, the viscosity is given by \cite{Skellandbook1967, BirdbookAH1987,
CarreaubookKC1997}

\begin{equation}
\mu=\frac{\tau_{o}}{\gamma}+C'
\end{equation}
where $\tau_{o}$ is the yield stress and $C'$ is the viscosity consistency coefficient. On applying
the variational principle, as formulated by Equation \ref{MainEq}, and following the previously
outlined method we obtain

\begin{equation}\label{BingGamR}
\tau_{o}\ln\gamma+C'\gamma=Az+D
\end{equation}
where $A$ and $D$ are the constants of integration. Using the boundary conditions at the center
plane and at the slit wall and following a similar procedure to that of Casson, we obtain

\begin{equation}
D=0 \hspace{1cm} {\rm and} \hspace{1cm} A=\frac{\tau_{o}}{B}\ln\left(\frac{B\Delta
p}{LC'}-\frac{\tau_{o}}{C'}\right)+\left(\frac{\Delta p}{L}-\frac{\tau_{o}}{B}\right)
\end{equation}
The strain rate is then obtained numerically from Equation \ref{BingGamR}, and thereby $v$ and $Q$
are computed successively, as explained before.

In Figure \ref{BinghamFig} two sample cases of the flow of Bingham fluids in thin slits are
presented. As seen, the agreement between the WRMS solutions, as obtained from Equation
\ref{QeqBing}, and the variational solutions are rather good despite the fact that the variational
method is an approximation when applied to viscoplastic fluids.


\begin{figure}
\centering %
\subfigure[]%
{\begin{minipage}[b]{0.5\textwidth} \CIF {g/BinghamFig1}
\end{minipage}}
\Hs
\subfigure[]%
{\begin{minipage}[b]{0.5\textwidth} \CIF {g/BinghamFig2}
\end{minipage}}
%
\caption{Comparing the WRMS analytical solutions, as given by Equation \ref{QeqBing}, to the
variational solutions for the flow of Bingham fluids in long thin slits with
(a) $C=0.02$~Pa.s, $\tau_o=0.25$~Pa, $B=0.015$~m, $W=1.0$~m and $L=0.75$~m;
and (b) $C=0.034$~Pa.s, $\tau_o=0.75$~Pa, $B=0.018$~m, $W=1.0$~m and $L=1.25$~m.
The axes are as in Figure \ref{ReeEyringFig}, while the differences are about 1.12\% and 1.96\%
respectively. \label{BinghamFig}}
\end{figure}

\subsection{Herschel-Bulkley}

The viscosity of Herschel-Bulkley fluids is given by \cite{Skellandbook1967, BirdbookAH1987,
CarreaubookKC1997}

\begin{equation}
\mu=\frac{\tau_{o}}{\gamma}+C\gamma^{n-1}
\end{equation}
where $\tau_{o}$ is the yield stress, $C$ is the viscosity consistency coefficient and $n$ is the
flow behavior index. On following a procedure similar to the procedure of Bingham model with the
application of the $\gamma$ two boundary conditions, we get the following equation

\begin{equation}\label{HBGamR}
\tau_{o}\ln\gamma+\frac{C}{n}\gamma^{n}=Az
\end{equation}
where

\begin{equation}
A=\frac{\tau_{o}}{B}\ln\left[\left(\frac{B\Delta
p}{LC}-\frac{\tau_{o}}{C}\right)^{1/n}\right]+\frac{1}{n}\left(\frac{\Delta
p}{L}-\frac{\tau_{o}}{B}\right)
\end{equation}
On solving Equation \ref{HBGamR} numerically, $\gamma$ as a function of $z$ is obtained, followed
by obtaining $v$ and $Q$, as explained previously.

\begin{figure}
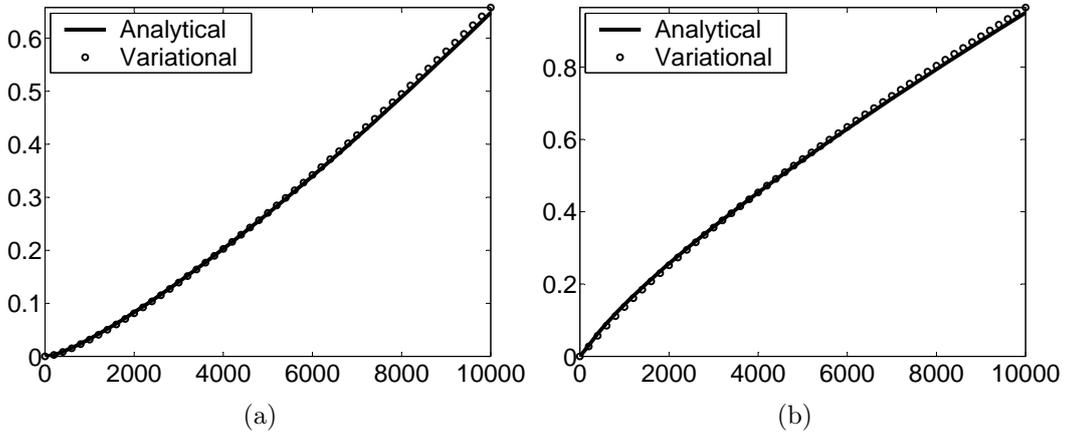

\centering %
\subfigure[]%
{\begin{minipage}[b]{0.5\textwidth} \CIF {g/HBFig1}
\end{minipage}}
\Hs
\subfigure[]%
{\begin{minipage}[b]{0.5\textwidth} \CIF {g/HBFig2}
\end{minipage}}
%
\caption{Comparing the WRMS analytical solutions, as given by Equation \ref{QeqHB}, to the
variational solutions for the flow of Herschel-Bulkley fluids in long thin slits with
(a) $n=0.8$, $C=0.05$~Pa.s$^n$, $\tau_o=0.5$~Pa, $B=0.01$~m, $W=1.0$~m and $L=1.2$~m;
and (b) $n=1.25$, $C=0.025$~Pa.s$^n$, $\tau_o=1.0$~Pa, $B=0.03$~m, $W=1.0$~m and $L=1.3$~m.
The axes are as in Figure \ref{ReeEyringFig}, while the differences are about 1.49\% and 1.74\%
respectively. \label{HBFig}}
\end{figure}

In Figure \ref{HBFig} we compare the WRMS analytical solutions of Equation \ref{QeqHB} to the
variational solutions for two sample Herschel-Bulkley fluids, one shear thinning and one shear
thickening, both with yield stress. As seen, the agreement between the solutions of the two methods
is good as in the previous cases of Casson and Bingham fluids.


\section{Conclusions} \label{Conclusions}

In this paper we provided further evidence for the validity of the variational method which is
based on minimizing the total stress in the flow conduit to obtain flow relations for the
generalized Newtonian fluids in confined geometries. Our investigation in the present paper, which
is related to the plane long thin slit geometry, confirms our previous findings which were
established using the straight circular uniform tube geometry. Eight fluid types are used in the
present investigation: Newtonian, power law, Ree-Eyring, Carreau, Cross, Casson, Bingham and
Herschel-Bulkley. This effort, added to the previous investigations, should be sufficient to
establish the variational method and the optimization principle upon which the method rests. For
the Newtonian and power law fluids, full analytical solutions are obtained from the variational
method, while for the other fluids mixed analytical-numerical procedures were established and used
to obtain the solutions.

Although some of the derived expressions and solutions are not of interest of their own as they can
be easily obtained from other non-variational methods, the theoretical aspect of our investigation
should be of great interest as it reveals a tendency of the flow system to minimize the total
stress which the variational method is based upon; hence giving an insight into the underlying
physical principles that control the flow of fluids.

The value of our investigation is not limited to the above mentioned theoretical aspect but it has
a practical aspect as well since the variational method can be used as an alternative to other
methods for other geometries and other rheological fluid models where mathematical difficulties may
be overcome in one formulation based on one of these methods but not the others. The variational
method is also more general and hence enjoys a wider applicability than some of the other methods
which are based on more special or restrictive physical or mathematical principles.

\clearpage
\section{Nomenclature}

\begin{supertabular}{ll}
$\gamma$                &   rate of shear strain (s$^{-1}$) \\
$\delta$                &   $\mu_{0}-\mu_{i}$ (Pa.s) \\
$\lambda$               &   characteristic time constant (s) \\
$\mu$                   &   fluid shear viscosity (Pa.s) \\
$\mu_{0}$               &   zero-shear viscosity (Pa.s) \\
$\mu_{i}$               &   infinite-shear viscosity (Pa.s) \\
$\mu_{o}$               &   Newtonian viscosity (Pa.s) \\
$\mu_{r}$               &   low-shear viscosity in Ree-Eyring model (Pa.s) \\
$\sigma_\parallel$      &   area of slit wall tangential to the flow direction (m$^2$) \\
$\tau$                  &   shear stress (Pa) \\
$\tau_{\pm B}$          &   shear stress at slit walls corresponding to $z=\pm B$ (Pa) \\
$\tau_c$                &   characteristic shear stress in Ree-Eyring model (Pa) \\
$\tau_o$                &   yield stress in Casson, Bingham and Herschel-Bulkley models (Pa) \\
$\tau_t$                &   total shear stress (Pa) \\
\\
$B$                     &   slit half thickness (m) \\
$C$                     &   viscosity consistency coefficient in Herschel-Bulkley model (Pa.s$^{n}$) \\
$C'$                    &   viscosity consistency coefficient in Bingham model (Pa.s) \\
$f$                     &   $\lambda^{m}\gamma_{B}^{m}$ \\
$_{2}F_{1}$             &   hypergeometric function \\
$F_\perp$               &   force normal to the slit cross section (N) \\
$g$                     &   $1+f$ \\
$I_{\rm Ca}$            &   definite integral expression for Carreau model (Pa$^2$.s$^{-1}$) \\
$I_{\rm Cr}$            &   definite integral expression for Cross model (Pa$^2$.s$^{-1}$) \\
$k$                     &   viscosity consistency coefficient in power law model (Pa.s$^n$) \\
$K$                     &   viscosity consistency coefficient in Casson model (Pa.s) \\
$L$                     &   slit length (m) \\
$m$                     &   indicial parameter in Cross model \\
$n$                     &   flow behavior index in power law, Carreau and Herschel-Bulkley models \\
$\Delta p$              &   pressure drop across the slit length (Pa) \\
$P_0$                   &   pressure at the slit entrance (Pa) \\
$P_L$                   &   pressure at the slit exit (Pa) \\
$Q$                     &   volumetric flow rate (m$^{3}$.s$^{-1}$) \\
$v$                     &   fluid speed in the flow direction (m.s$^{-1}$) \\
$W$                     &   slit width (m) \\
$z$                     &   coordinate of slit smallest dimension (m) \\
\end{supertabular}

\clearpage
\phantomsection \addcontentsline{toc}{section}{References} %
\bibliographystyle{unsrt}

\clearpage
\section{Appendix}\label{Appendix}

Here we derive a general formula for the volumetric flow rate of generalized Newtonian fluids in
rigid plane long thin uniform slits following a method similar to the one ascribed to Weissenberg,
Rabinowitsch, Mooney, and Schofield \cite{Skellandbook1967}, and hence we label the method with
WRMS. We then apply it to obtain analytical relations correlating the volumetric flow rate to the
pressure drop for the flow of Newtonian and seven non-Newtonian fluids through the above described
slit geometry.

The differential flow rate through a differential strip along the slit width is given by

\begin{equation}
dQ=vWdz
\end{equation}
where $Q$ is the volumetric flow rate and $v\equiv v(z)$ is the fluid speed at $z$ in the $x$
direction according to the coordinates system demonstrated in Figure \ref{SlitPlot}. Hence

\begin{equation}
\frac{Q}{W}=\int_{-B}^{+B}vdz
\end{equation}
On integrating by parts we get

\begin{equation}
\frac{Q}{W}=\left[vz\right]_{-B}^{+B}-\int_{v_{-B}}^{v_{+B}}zdv
\end{equation}
The first term on the right hand side is zero due to the no-slip boundary condition, and hence we
have

\begin{equation}
\frac{Q}{W}=-\int_{v_{-B}}^{v_{+B}}zdv\label{eqQ}
\end{equation}
Now, from Equation \ref{tauBeq}, we have

\begin{equation}
\tau_{\pm B}=\frac{B\Delta p}{L}
\end{equation}
where $\tau_{\pm B}$ is the shear stress at the slit walls. Similarly we have

\begin{equation}
\tau_{z}=\frac{z\Delta p}{L}
\end{equation}
where $\tau_{z}$ is the shear stress at $z$. Hence

\begin{equation}
\tau_{z}=\frac{z}{B}\tau_{\pm B}\,\,\,\,\,\,\,\,\Rightarrow\,\,\,\,\,\,\,\,\,
z=\frac{B\tau_{z}}{\tau_{\pm B}}\,\,\,\,\,\,\,\,\Rightarrow\,\,\,\,\,\,\,\,\,
dz=\frac{Bd\tau_{z}}{\tau_{\pm B}}\label{eqAAA}
\end{equation}
We also have

\begin{equation}
\gamma_z=-\frac{dv}{dz}\,\,\,\,\,\,\,\,\Rightarrow\,\,\,\,\,\,\,\,\, dv=-\gamma_z
dz=-\gamma_z\frac{Bd\tau_{z}}{\tau_{\pm B}}\label{eqBBB}
\end{equation}
Now due to the symmetry with respect to the plane $z=0$ we have

\begin{equation}
\tau_{B}\equiv\tau_{+B}=\tau_{-B}
\end{equation}
On substituting from Equations \ref{eqAAA} and \ref{eqBBB} into Equation \ref{eqQ}, considering the
flow symmetry across the center plane $z=0$, and changing the limits of integration we obtain

\begin{equation}
\frac{Q}{W}=\int_{\tau_{-B}}^{\tau_{+B}}\frac{B\tau_{z}}{\tau_{\pm
B}}\gamma_z\frac{Bd\tau_{z}}{\tau_{\pm
B}}=2\left(\frac{B}{\tau_{B}}\right)^{2}\int_{0}^{\tau_{B}}\tau_{z}\gamma_z d\tau_{z}
\end{equation}
that is

\begin{equation}
\boxed{ Q=2W\left(\frac{B}{\tau_{B}}\right)^{2}\int_{0}^{\tau_{B}}\gamma \tau d\tau }
\end{equation}
where it is understood that $\gamma=\gamma_{z}\equiv\gamma(z)$ and $\tau=\tau_{z}\equiv\tau(z)$.

\vspace{0.5cm}

For \textbf{Newtonian} fluids with viscosity $\mu_{o}$ we have

\begin{equation}
\tau=\mu_{o}\gamma\,\,\,\,\,\,\,\,\,\Rightarrow\,\,\,\,\,\,\,\,\,\gamma=\frac{\tau}{\mu_{o}}
\end{equation}
Hence

\begin{equation}
Q=2W\left(\frac{B}{\tau_{B}}\right)^{2}\int_{0}^{\tau_{B}}\gamma\tau
d\tau=\frac{2W}{\mu_{o}}\left(\frac{B}{\tau_{B}}\right)^{2}\int_{0}^{\tau_{B}}\tau^{2}\,d\tau
=\frac{2WB^{2}}{3\mu_{o}}\tau_{B}
\end{equation}
that is

\begin{equation}\label{QeqNewt}
\boxed{ Q=\frac{2WB^{3}\Delta p}{3\mu_{o}L} }
\end{equation}

\vspace{0.5cm}

For \textbf{power law} fluids we have

\begin{equation}
\tau=k\gamma^{n}\,\,\,\,\,\,\,\,\,\,\,\,\Rightarrow\,\,\,\,\,\,\,\,\,\gamma=\left(\frac{\tau}{k}\right)^{1/n}
\end{equation}
Hence

\begin{equation}
Q=2W\left(\frac{B}{\tau_{B}}\right)^{2}\int_{0}^{\tau_{B}}\left(\frac{\tau}{k}\right)^{1/n}\tau
d\tau=\frac{2W}{k^{1/n}}\left(\frac{B}{\tau_{B}}\right)^{2}\int_{0}^{\tau_{B}}\tau^{1+1/n}\,d\tau
\end{equation}

\begin{equation}
Q=\frac{2W}{k^{1/n}\left(2+1/n\right)}\left(\frac{B}{\tau_{B}}\right)^{2}\tau_{B}^{2+1/n}=\frac{2WB^{2}}{k^{1/n}\left(2+1/n\right)}\tau_{B}^{1/n}
\end{equation}
that is

\begin{equation}\label{QeqPl}
\boxed{ Q=\frac{2WB^{2}n}{2n+1}\sqrt[n]{\frac{B\Delta p}{kL}} }
\end{equation}
When $n=1$, with $k\equiv\mu_{o}$, the formula reduces to the Newtonian, Equation \ref{QeqNewt}, as
it should be.

\vspace{0.5cm}

For \textbf{Ree-Eyring} fluids we have

\begin{equation}
\tau=\tau_{c}\,\mathrm{asinh}\left(\frac{\mu_{r}\gamma}{\tau_{c}}\right)\,\,\,\,\,\,\,\,\Rightarrow\,\,\,\,\,\,\,\,\gamma=\frac{\tau_{c}}{\mu_{r}}\sinh\left(\frac{\tau}{\tau_{c}}\right)
\end{equation}
Hence

\begin{equation}
Q=\frac{2W\tau_{c}}{\mu_{r}}\left(\frac{B}{\tau_{B}}\right)^{2}\int_{0}^{\tau_{B}}\tau\sinh\left(\frac{\tau}{\tau_{c}}\right)d\tau
\end{equation}

\begin{equation}
Q=\frac{2W\tau_{c}^{2}}{\mu_{r}}\left(\frac{B}{\tau_{B}}\right)^{2}\left[\tau\,\cosh\left(\frac{\tau}{\tau_{c}}\right)-\tau_{c}\,\sinh\left(\frac{\tau}{\tau_{c}}\right)\right]_{0}^{\tau_{B}}
\end{equation}
that is

\begin{equation}\label{QeqRE}
\boxed{
Q=\frac{2W\tau_{c}^{2}}{\mu_{r}}\left(\frac{B}{\tau_{B}}\right)^{2}\left[\tau_{B}\,\cosh\left(\frac{\tau_{B}}{\tau_{c}}\right)-\tau_{c}\,\sinh\left(\frac{\tau_{B}}{\tau_{c}}\right)\right]
}
\end{equation}

\vspace{0.5cm}

For \textbf{Carreau} fluids, the viscosity is given by

\begin{equation}
\mu=\frac{\tau}{\gamma}=\mu_{i}+\delta\left(1+\lambda^{2}\gamma^{2}\right)^{n'/2}
\end{equation}
where $\delta=\left(\mu_{0}-\mu_{i}\right)$ and $n'=\left(n-1\right)$. Therefore

\begin{equation}
\tau=\gamma\left[\mu_{i}+\delta\left(1+\lambda^{2}\gamma^{2}\right)^{n'/2}\right]\label{tauEqCar}
\end{equation}
and hence

\begin{equation}
d\tau=\left[\mu_{i}+\delta\left(1+\lambda^{2}\gamma^{2}\right)^{n'/2}+n'\delta\lambda^{2}\gamma^{2}\left(1+\lambda^{2}\gamma^{2}\right)^{(n'-2)/2}\right]d\gamma\label{eqdtauCar}
\end{equation}
Now, from the WRMS method we have

\begin{equation}
\frac{Q\,\tau_{B}^{2}}{2WB^{2}}=\int_{0}^{\tau_{B}}\gamma\tau d\tau\label{eqWRMSSl}
\end{equation}
If we label the integral on the right hand side of Equation \ref{eqWRMSSl} with $I_{\rm Ca}$ and
substitute for $\tau$ from Equation \ref{tauEqCar}, substitute for $d\tau$ from Equation
\ref{eqdtauCar}, and change the integration limits we obtain

\begin{equation}
I_{\rm
Ca}=\int_{0}^{\gamma_{B}}\gamma^{2}\left[\mu_{i}+\delta\left(1+\lambda^{2}\gamma^{2}\right)^{n'/2}\right]\left[\mu_{i}+\delta\left(1+\lambda^{2}\gamma^{2}\right)^{n'/2}+n'\delta\lambda^{2}\gamma^{2}\left(1+\lambda^{2}\gamma^{2}\right)^{(n'-2)/2}\right]d\gamma\label{eqICarSl2}
\end{equation}
On solving this integral equation analytically and evaluating it at its two limits we obtain

\begin{eqnarray}
I_{\rm Ca} & = & \frac{n'\delta^{2}\gamma_{B}\left[_{2}F_{1}\left(\frac{1}{2},1-n';\frac{3}{2};-\lambda^{2}\gamma_{B}^{2}\right)-{}_{2}F_{1}\left(\frac{1}{2},-n';\frac{3}{2};-\lambda^{2}\gamma_{B}^{2}\right)\right]}{\lambda^{2}} \nonumber \\
 & + & \frac{\left(1+n'\right)\delta^{2}\gamma_{B}^{3}\,{}_{2}F_{1}\left(\frac{3}{2},-n';\frac{5}{2};-\lambda^{2}\gamma_{B}^{2}\right)}{3} \nonumber \\
 & + & \frac{n'\delta\mu_{i}\gamma_{B}\left[_{2}F_{1}\left(\frac{1}{2},1-\frac{n'}{2};\frac{3}{2};-\lambda^{2}\gamma_{B}^{2}\right)-{}_{2}F_{1}\left(\frac{1}{2},-\frac{n'}{2};\frac{3}{2};-\lambda^{2}\gamma_{B}^{2}\right)\right]}{\lambda^{2}} \nonumber \\
 & + & \frac{\left(2+n'\right)\delta\mu_{i}\gamma_{B}^{3}\,{}_{2}F_{1}\left(\frac{3}{2},-\frac{n'}{2};\frac{5}{2};-\lambda^{2}\gamma_{B}^{2}\right)+\mu_{i}^{2}\gamma_{B}^{3}}{3}
\end{eqnarray}
where $_{2}F_{1}$ is the hypergeometric function of the given arguments with its real part being
used in this evaluation. Now, from applying the rheological equation at the slit wall we have

\begin{equation}
\left[\mu_{i}+\delta\left(1+\lambda^{2}\gamma_{B}^{2}\right)^{n'/2}\right]\gamma_{B}=\frac{B\,\Delta
p}{L}
\end{equation}
From the last equation, $\gamma_{B}$ can be obtained numerically by a simple numerical solver, such
as bisection, and hence $I_{\rm Ca}$ is computed. The volumetric flow rate is then obtained from

\begin{equation}\label{QeqCar}
\boxed{ Q=\frac{2WB^{2}I_{\rm Ca}}{\tau_{B}^{2}} }
\end{equation}

\vspace{0.5cm}

For \textbf{Cross} fluids, the viscosity is given by

\begin{equation}
\mu=\frac{\tau}{\gamma}=\mu_{i}+\frac{\delta}{1+\lambda^{m}\gamma^{m}}
\end{equation}
where $\delta=\left(\mu_{0}-\mu_{i}\right)$. Therefore

\begin{equation}
\tau=\gamma\left(\mu_{i}+\frac{\delta}{1+\lambda^{m}\gamma^{m}}\right)\label{tauEqCro}
\end{equation}
and hence

\begin{equation}
d\tau=\left(\mu_{i}+\frac{\delta}{1+\lambda^{m}\gamma^{m}}-\frac{m\delta\lambda^{m}\gamma^{m}}{\left(1+\lambda^{m}\gamma^{m}\right)^{2}}\right)d\gamma\label{eqdtauCro}
\end{equation}
If we follow a similar procedure to that of Carreau by applying the WRMS method and labeling the
right hand side integral with $I_{\rm Cr}$, substituting for $\tau$ and $d\tau$ from Equations
\ref{tauEqCro} and \ref{eqdtauCro} respectively, and changing the integration limits of $I_{\rm
Cr}$ we get

\begin{equation}
I_{\rm
Cr}=\int_{0}^{\gamma_{B}}\gamma^{2}\left(\mu_{i}+\frac{\delta}{1+\lambda^{m}\gamma^{m}}\right)\left(\mu_{i}+\frac{\delta}{1+\lambda^{m}\gamma^{m}}-\frac{m\delta\lambda^{m}\gamma^{m}}{\left(1+\lambda^{m}\gamma^{m}\right)^{2}}\right)d\gamma\label{eqICroSl2}
\end{equation}
On solving this integral equation analytically and evaluating it at its two limits we obtain

\begin{equation}
I_{\rm Cr}=\frac{\left[3\delta^{2}\left(m-g\right)-\left\{
\delta^{2}\left(m-3\right)+2m\delta\mu_{i}\right\}
g^{2}\,_{2}F_{1}\left(1,\frac{3}{m};1+\frac{3}{m};-f\right)+6m\delta\mu_{i}g+2m\mu_{i}^{2}g^{2}\right]\gamma_{B}^{3}}{6mg^{2}}
\end{equation}
where

\begin{equation}
f=\lambda^{m}\gamma_{B}^{m}\,\,\,\,\,\,\,\,\,\,\,\,,\,\,\,\,\,\,\,\,\,\,\,\, g=1+f\end{equation}
and $_{2}F_{1}$ is the hypergeometric function of the given argument with its real part being used
in this evaluation. As before, from applying the rheological equation at the wall we have

\begin{equation}
\left(\mu_{i}+\frac{\delta}{1+\lambda^{m}\gamma_{B}^{m}}\right)\gamma_{B}=\frac{B\,\Delta
p}{L}\end{equation}
From this equation, $\gamma_{B}$ can be obtained numerically and hence $I_{\rm Cr}$ is computed.
Finally, the volumetric flow rate is obtained from

\begin{equation}\label{QeqCro}
\boxed{ Q=\frac{2WB^{2}I_{\rm Cr}}{\tau_{B}^{2}} }
\end{equation}

\vspace{0.5cm}

For \textbf{Casson} fluids we have

\begin{equation}
\tau^{1/2}=\left(K\gamma\right)^{1/2}+\tau_{o}^{1/2}\hspace{2cm}(\tau\ge\tau_{o})
\end{equation}
Hence

\begin{equation}
\gamma=\frac{\left(\tau^{1/2}-\tau_{o}^{1/2}\right)^{2}}{K}
\end{equation}
On applying the WRMS equation we get

\begin{equation}
Q=\frac{2W}{K}\left(\frac{B}{\tau_{B}}\right)^{2}\int_{\tau_{o}}^{\tau_{B}}\tau\left(\tau^{1/2}-\tau_{o}^{1/2}\right)^{2}d\tau
\end{equation}

\begin{equation}
Q=\frac{2W}{K}\left(\frac{B}{\tau_{B}}\right)^{2}\int_{\tau_{o}}^{\tau_{B}}\left(\tau^{2}-2\sqrt{\tau_{o}}\tau^{3/2}+\tau_{o}\tau\right)d\tau
\end{equation}

\begin{equation}
Q=\frac{2W}{K}\left(\frac{B}{\tau_{B}}\right)^{2}\left[\frac{\tau^{3}}{3}-\frac{4\sqrt{\tau_{o}}\tau^{5/2}}{5}+\frac{\tau_{o}\tau^{2}}{2}\right]_{\tau_{o}}^{\tau_{B}}
\end{equation}
that is

\begin{equation}\label{QeqCas}
\boxed{
Q=\frac{2W}{K}\left(\frac{B}{\tau_{B}}\right)^{2}\left[\frac{\tau_{B}^{3}}{3}-\frac{4\sqrt{\tau_{o}}\tau_{B}^{5/2}}{5}+\frac{\tau_{o}\tau_{B}^{2}}{2}-\frac{\tau_{o}^{3}}{30}\right]
}
\end{equation}

\vspace{0.5cm}

For \textbf{Bingham} fluids we have

\begin{equation}
\tau=\tau_{o}+C'\gamma\,\,\,\,\,\,\,\,\,\,\,\,\Rightarrow\,\,\,\,\,\,\,\,\,\,\,\,\gamma=\frac{\tau-\tau_{o}}{C'}\hspace{2cm}(\tau\ge\tau_{o})
\end{equation}
Hence

\begin{equation}
Q=\frac{2W}{C'}\left(\frac{B}{\tau_{B}}\right)^{2}\int_{\tau_{o}}^{\tau_{B}}\tau\left(\tau-\tau_{o}\right)d\tau=\frac{2W}{C'}\left(\frac{B}{\tau_{B}}\right)^{2}\left[\frac{\tau^{3}}{3}-\frac{\tau_{o}\tau^{2}}{2}\right]_{\tau_{o}}^{\tau_{B}}
\end{equation}
that is

\begin{equation}\label{QeqBing}
\boxed{
Q=\frac{2W}{C'}\left(\frac{B}{\tau_{B}}\right)^{2}\left[\frac{\tau_{B}^{3}}{3}-\frac{\tau_{o}\tau_{B}^{2}}{2}+\frac{\tau_{o}^{3}}{6}\right]
}
\end{equation}
When $\tau_{o}=0$, with $C'\equiv\mu_{o}$, the formula reduces to the Newtonian, Equation
\ref{QeqNewt}, as it should be.

\vspace{0.5cm}

For \textbf{Herschel-Bulkley} fluids we have

\begin{equation}
\tau=\tau_{o}+C\gamma^{n}\,\,\,\,\,\,\,\,\,\,\,\,\Rightarrow\,\,\,\,\,\,\,\,\,\,\,\,\gamma=\frac{1}{C^{1/n}}\left(\tau-\tau_{o}\right)^{1/n}\hspace{2cm}(\tau\ge\tau_{o})
\end{equation}
Hence

\begin{equation}
Q=\frac{2W}{C^{1/n}}\left(\frac{B}{\tau_{B}}\right)^{2}\int_{\tau_{o}}^{\tau_{B}}\tau\left(\tau-\tau_{o}\right)^{1/n}d\tau
\end{equation}

\begin{equation}
Q=\frac{2W}{C^{1/n}}\left(\frac{B}{\tau_{B}}\right)^{2}\left[\frac{n\left(n\tau_{o}+n\tau+\tau\right)\left(\tau-\tau_{o}\right)^{1+1/n}}{\left(2n^{2}+3n+1\right)}\right]_{\tau_{o}}^{\tau_{B}}
\end{equation}
that is

\begin{equation}\label{QeqHB}
\boxed{
Q=\frac{2W}{C^{1/n}}\left(\frac{B}{\tau_{B}}\right)^{2}\left[\frac{n\left(n\tau_{o}+n\tau_{B}+\tau_{B}\right)\left(\tau_{B}-\tau_{o}\right)^{1+1/n}}{\left(2n^{2}+3n+1\right)}\right]
}
\end{equation}
When $n=1$, with $C\equiv C'$, the formula reduces to the Bingham, Equation \ref{QeqBing}; when
$\tau_{o}=0$, with $C\equiv k$, the formula reduces to the power law, Equation \ref{QeqPl}; and
when $n=1$ and $\tau_{o}=0$, with $C\equiv\mu_{o}$, the formula reduces to the Newtonian, Equation
\ref{QeqNewt}, as it should be.

\end{document}